\documentstyle[12pt]{article}
\font\mygotic eufm10 scaled 1200

\title{Derivation of Chiral Lagrangians from Random Lattice QCD}

\author{ Oleg V. Pavlovsky \thanks{e-mail address:
ovp@goa.bog.msu.su }  \\
{\em
Institute for Theoretical Problems of Microphysics,} \\
{\em Moscow State University } \\ {\em Moscow, 119992, Russian
Federation.} }

\date{ \ \ \  }

\begin{document}

\maketitle

\begin{abstract}
In our work we extend the ideas of the derivation of the chiral effective theory
from the lattice QCD [1] to the case of the random lattice regularization of QCD.
Such procedure allows in principle to find
contribution of any order into the chiral effective lagrangian.
It is shown that an infinite subseries of the chiral perturbation can be
summed up into tne Born-Infeld term and the logarithmic correction to
them.
\end{abstract}

\vspace{1cm}

PACS number(s): 11.15.Ha, 12.38.Gc, 12.39.Fe, 11.15.Me, 11.30.Cp,
11.30.Rd, 12.39.Dc

\section{Introduction and motivation}

At low energies strong interactions can be described by
effective chiral lagragians. Such effective field approach is very
important, both for the physics of the hadron-hadron interaction
and for the concept of the low-energy baryon state.
Therefore, there is a strong motivation to derive effective chiral theories from
QCD.

The first essential contribution
to this topic was made in~\cite{witch}.  Using the method of
large $N$ expansion,  an effective chiral theory was suggested in the
form of a series of chiral invariants.
The second order theory
of such type is the well-known Skyrme model \cite{skyrme} of
low-energy baryon states, a phenomenological unified theory for
mesons and baryons where the baryon is treated as a topological soliton
of nonlinear chiral fields.  Another interesting approach was
proposed in \cite{novozh} where the Skyrme model was derived from
integration of the chiral anomaly.

All such methods play a very essential role in particle physics and
give appropriate description of the behavior of the chiral field at
low energy.
However, to study the chiral field near
the confinement boundary one must analyze the whole series of chiral
perturbations.  The Chiral Bag Model \cite{bag} provides a
very good illustration of this problem.  In this model the boundary
between chiral fields and color fields is specified by hand.  Although
this gives quite good agreement with experimental data,
it is not clear what might be the physical
mechanism of formation of this chiral bag.

Fortunately, Lattice QCD numerical experiments give us today a lot of
interesting information about the behavior of color fields
(quarks and gluons) in the strong coupling regime.  These data are
very essential to understand the physics of the baryon at low energies;
in fact the lattice approach has formed empirical and theoretical basis of
the baryon string model~\cite{barstring}.

Against the background of these facts, it looks astonishing that the
Skyrme model should give a good agreement with the experiment data without
accounting for the color degrees of freedom at all.  Therefore the
question of unification of these paradigms: the chiral soliton model and the
model of baryon string, becomes very essential.
It is natural to begin the search for the way of such unification
with analyzing the chiral limit of QCD on the lattice.

\section{Why do we need the Random Lattice QCD}

Derivation of a chiral effective Lagrangian from lattice
QCD has been attempted many times since long ago.  The
well-known Brezin \& Gross trick \cite{Brezin} makes it possible to
perform integration of the link matrix in the strong coupling regime and
to obtain various first order chiral effective theories
\cite{forder}.

Although at first such approaches led to great success, they have not been very
popular, because they do not allow to obtain any
corrections to first order results.  Lattice regularization
breaks the rotational symmetry of the initial theory from the
continuous rotation group down to a discrete group of rotations at fixed
angles.  Hence, lattice regularization approaches give correct
results only for those tensors that are invariant with respect to such
discrete groups.  In particular, using the ordinary Hyper-Cubical (HC)
lattice, one can obtain only a first order effective theory, while
for corrections this method generates non-rotation invariant (non-Lorentz
invariant) terms. Generation of high-order effective field
theories requires a more symmetrical lattice.

The problem of breakdown of rotational symmetry on a lattice has
been attracting important attention for a long time.  It was shown
\cite{f4} that in 4 dimensions the so-called  Body Centered Hyper-Cubical
(BCHC) or F4 lattice has the largest discrete symmetry  group.  (BCHC
consists from the all sites of the HC lattice together with centers
of its elementary cells.)
This property of the BCHC lattice gives a
possibility to obtain the next-to-leading (NL) correction to the
first order of the chiral perturbation theory \cite{rebbi}.

The results
of the papers \cite{rebbi} are essential for our analysis, as
they confirm the effectiveness of
the idea of chiral effective lagrangian derivation from the lattice QCD.
Moreover, these
results are interesting from phenomenological point of view
because, as is well known \cite{skyrme}, the NL~corrections violate
the scale invariance of the prototype (first order) chiral theory
that leads to generation of chiral topological solitons
(Skyrmions).  The Next-Leading order chiral effective theory that was
implemented in \cite{rebbi} is in agreement with our phenomenological
propositions  \cite{Gasser},
and in our work we will use
methodological ideas from \cite{rebbi} in order to define the
behaviour of chiral field near the confinement surface.


As one could see, in order to solve our problem, the Next-Leading
order corrections are not enough. This theory has no solutions that
look like chiral ``bag". Moreover, as will be shown later, near the
confinement surface (near the source of the chiral field), the influence
of high order corrections became larger and larger. But the BCHC
lattice method gives the NL corrections only and the further use of
this method for the defining of the high order terms leads to
generation of non-relativistic (non-rotational) invariant terms. It
means that we need a more symmetric lattice than the BCHC lattice.

Unfortunately, a lattice which would be more symmetric than BCHC
lattice cannot be constructed in 4 dimension. Moreover any method
based on a lattice of a fixed geometry has artifacts
coming from priority directions that correspond to basis
vectors of the lattice. It is these artifacts that eventually lead to the
problems with the rotational (relativistic) invariance rendering the use
of the BCHC lattice to be only half measure. For
solving our problem a modification of the initial concept of
lattice regularization must be performed. We need to find a
concept of lattice regularization that has no priority
directions. Fortunately this concept is known for a long time and
is called the Random Lattice approach \cite{RL}.

The idea of  Random Lattice was proposed originally by Voronoi and
Delaunay: today this method is widely used in the modern science.
For the quantum field theory the method was modified by
Christ, Friedberg and Lee \cite{RL}. In these articles it has been
shown that in order to restor the Lorentz
(rotational) invariance, it is necessary to perform an average over
an ensemble of random lattices. As a result one gets the averaging over
all possible directions and it is intuitively clear that this
procedure leads to the disappearance of the artifacts that cause the
violation of the group of the space rotations.

But how to perform such random discretization? This procedure has
the tree steps:

1) Pick $N$ sites $x_i$ at random in the volume $V$.

2) Associate with each $x_i$ a so-called Voronoi cell $c_i$
$$
c_i = \{ x \mid d(x,x_i) \leq d(x,x_j), \forall j \neq i  \}
$$
where $d(x,y)$ is a distance between points $x$ and $y$. It
means  that the Voronoi cell $c_i$ consists of all points $x$ that are
closer to the center site $x_i$ than to any other site.

3) Constrict the dual Delaunay lattice by linking the center sites
of all Voronoi cells which share a common face.

Now if one considers the the big ensemble of such
Voronoi-Delaunay random lattices based on  various distributions of
sites $x_i$, it possible to prove that the original rotational symmetry
is restored \cite{RL}. In our work we use this procedure to
obtain an effective chiral lagrangian from lattice QCD. This
methodological point of view it is a modification of the method
proposed in \cite{rebbi} in the case of the Random Lattice approach.

\section{From Lattice QCD to chiral lagrangians: step by step}

Now let me briefly recall a general steps of the algorithm
 of derivation of the chiral lagrangian from the lattice QCD that
 was proposed in \cite{rebbi}.

{\bf Step 1: Definitions}

The starting point of our analysis is a standard
lattice action with Willson fermions
$$
Z= \int [DG] [D \bar{\psi} ] [D \psi ] \exp \{-S_{\mbox{pl}}(G)
-S_q(G,\bar{\psi},\psi) - S_J \}
$$
where:

1) the plaquette gauge field term is defined by
$$
S_{\mbox{pl}}= \frac{2N_{c}}{g^2} \sum_{pl} \left[1 - \frac{1}{N_c}
Re G_{x,\mu} G_{x-\mu,\nu} G^+_{x+\nu,\mu} G^+_{x,\nu} \right], \,
\, G_{x,\mu}=\exp\{ig\int_{\mbox{link}} dx'_\mu {\cal{A}}_\mu ( x' )
\};
$$

2)the link fermions term is defined by
$$
S_q = \sum_{x,\mu} \mbox{tr} (\bar{A}_\mu (x) G_\mu (x) + G^+_\mu (x) A_\mu
(x) )
$$
$$
A_\mu(x)^a_b = \bar{\psi}_b(x+\mu) P^+_\mu \psi^a (x), \, \, \,
\bar{A}_\mu(x)^a_b = \bar{\psi}_b(x) P^-_\mu \psi^a (x+\mu)
$$
and $P^\pm_\mu=\frac{1}{2}(r \pm \gamma_\mu)$;

3) the source term is defined by
$$
S_J= \sum_x J^\alpha_\beta (x) M^\beta_\alpha (x), \, \, \,
M^\beta_\alpha = \frac{1}{N_c} \psi^{a,\beta}(x) \bar{\psi}_{a,
\alpha} (x).
$$

{\bf Step 2: Strong-coupling regime on the lattice and integration over the gauge field}

In order to realize the strong-coupling regime on the lattice let us
consider the limit of the large coupling constant $g$ ($g \to
\infty$). This limit was widely studied \cite{forder} and the main
result is that in such limit integration over the gauge field can be
performed. (Of course, the direct  integration is difficult since
there exists the plaquette term $S_{\mbox{pl}}$, but due to the
strong-coupling limit on the first step plaquette contributions are negligible with respect
to the contribution from the link
integral $S_{q}$.  The plaquette contributions could be considered
in the systematic manner as perturbations in $1/g$ \cite{forder}.)

Let us consider the leading order contribution in this
strong-coupling expansion. The integrals over the gauge degrees of
freedom can be calculated into the large $N$ limit by using  the
standard procedure \cite{forder} and the result of these calculations is
the following
\begin{equation}
Z= \int [D \bar{\psi} ] [D \psi ] \exp \{-N \sum_{x,\nu} \mbox{tr}
[F(\lambda(x,\nu))] - S_J \} , \label{z1}
\end{equation}
where $\lambda_\nu = - M(x)P^-_\nu M(x+\nu) P^+_\nu $ and
$$
F(\lambda) = \mbox{tr} [ (1 - \sqrt{1-\lambda}) ] - \mbox{tr} [
\log(1 - \frac{1}{2}\sqrt{1-\lambda})  ] .
$$

Interestingly, the function
$F(\lambda)$ has the typical form of the Born-Infeld action with
a first logarithmic correction. This is no coincidence. In \cite{IIB},
it was shown by means of very similar technique that the
low-energy theory of the IIB superstring has a Born-Infeld form.
From the methodological point of view we perform a similar analysis
for QCD on the lattice and it is important to note before starting
 our  proof that our result will have a Born-Infeld form too.

{\bf Step 3: Integration over the fermion field and chiral limit}

Our next step is the integration over the fermion degrees of freedom
in (\ref{z1}). Using the source technique it was shown \cite{rebbi}
that integral (\ref{z1}) can be re-written into the form of an
integral over the unitary boson matrix $M_x$
\begin{equation}
 Z=\int D M \exp S_{\mbox{eff}}(M).
 \label{z2}
\end{equation}

As a matter of principle, we already performed the transformation from
the color lattice degrees of freedom ($G$ and $\psi$) to the boson
lattice degrees of freedom ($M$). Now our task is to realize the
continuum limit of expression (\ref{z2}).

This step of our analysis amounts to studying of the
stationary points of the lattice action $S_{\mbox{eff}}$.
Fortunately this is a very well studied task \cite{gw}. This problem
is connected with well-known investigations of the critical behavior
of the chiral field on the lattice and with the problem of the phase
transformation on the lattice (for references see issue
\cite{Rossi}). In \cite{rebbi}, it was shown that for our task the
stationary point is
$$
\hat{M}_0 = u_0 \hat{I}, \, \, u_0 (m_q=0, r=1)  = 1/4.
$$
Now one can expressed $M(x)$ in terms of the pseudoscalar Goldstone
bosons
$$
M=u_0 \exp( i \pi_i \tau_i \gamma_5/f_\pi )= u_0 [ U(x)
\frac{1+\gamma_5}{2} + U^{+}(x) \frac{1-\gamma_5}{2}]
$$
and the effective action is given in the form of the Taylor
expansion around this stationary point
\begin{equation}
S_{\mbox{eff}} (U) = - N \sum_{k=1}^\infty
\frac{F^{(k)}(\lambda_0)}{k!} \sum_{x,\nu} \mbox{tr} [
(\lambda_\nu(x) - \lambda_0)^k ] \label{seff} .
\end{equation}

Let us consider the expansion of the chiral field $U=\exp(i \pi_i
\tau_i /f_\pi)$ on the lattice around the point $x$ in power of the
small step of the lattice $a$
$$
U(x+\nu)= U(x) + a (\partial_\nu U(x) ) +
\frac{a^2}{2}(\partial^2_\nu U(x) ) + \cdots .
$$
And for components of the Taylor expansion (\ref{seff}) one obtain
\begin{equation}
\begin{array}{ccrccc}
\mbox{tr}[(\lambda_\nu(x) - \lambda_0)]  &=& -2 \lambda_0 \mbox{tr}(\alpha)& & & \\
\mbox{tr}[(\lambda_\nu(x) - \lambda_0)^2]&=&  2 \lambda^2_0 \mbox{tr}(\alpha^2) & - 4 \lambda^2_0 \mbox{tr}(\alpha)  & &  \\
\mbox{tr}[(\lambda_\nu(x) - \lambda_0)^3]&=& -2 \lambda^3_0 \mbox{tr}(\alpha^3) & + 6 \lambda^3_0 \mbox{tr}(\alpha^2) & &  \\
\mbox{tr}[(\lambda_\nu(x) - \lambda_0)^4]&=&  2 \lambda^4_0
\mbox{tr}(\alpha^4) & - 8 \lambda^4_0 \mbox{tr}(\alpha^3)  &
+ 4 \lambda^4_0 \mbox{tr}(\alpha^2)&  \\
\mbox{tr}[(\lambda_\nu(x) - \lambda_0)^5]&=& -2 \lambda^5_0
\mbox{tr}(\alpha^5)& \cdots\cdots & \cdots\cdots & \cdots\cdots
\end{array} ,
\label{comp}
\end{equation} where $\alpha = a^2 \partial_\nu U \partial_\nu U^+ +
O(a^4)$.

{\bf Step 4: Problem of rotational symmetry violation: examples of the
Hyper-Cubical and Body Centered Hyper-Cubical lattices}

Expressions (\ref{comp}) are very essential because these are a
simplest illustration of all aspects of the violation of the
rotational symmetry on the lattice. For this moment we assume
nothing special about the structure of our lattice. We try to formulate our
result as generally as possible and all information about the lattice
contained in the vectors $\nu$ that correspond to the basic
vectors of the lattice (for example, the vectors $\nu$ for the
Hyper-Cubical lattice are the Cartesian basic vectors $\vec i$,
$\vec j$, $\vec k$ and $\vec t$). The leading order part can be
calculated trivially. Indeed, using the simple Hyper-Cubical lattice
where $\nu=i,j \, :$ $i=(1 \ldots 4)$ it is easy to show that the
leading order contribution is the prototype chiral lagrangian
\begin{equation}
 P_{O(p^2)} \sim \mbox{tr} [ \partial_\mu U
\partial^\mu U^+ ] .
\label{lorder}
\end{equation}

As I said before the rotational symmetry violation argument does
not allow to use the HC lattice calculation for the next-leading
order contributions.  For obtaining of these contributions a more
symmetrical lattice must be used. In \cite{f4} it was shown that
this lattice is a Body Centered Hyper-Cubical (BCHC) or F4 lattice.

Unfortunately, this method can not be directly
used for finding next contributions and the origin of this fact is
again the violation of the rotation symmetry  but now on the  F4
lattice. Moreover, there are no any more symmetrical lattice with
fixed positions of sites in 4-dimensional \cite{f4}. It means that we
need an absolutely different lattice concept that
guarantees the restoration of the initial symmetries. Fortunately
this concept is known now. This is the Random Lattice concept
(RL) \cite{RL}.

\section{Random Lattice in action}

The basic idea of the RL is the averaging over the big ensemble of
various lattices with random distributions of sites and it is possible
to show that such averaging leads to the restoration of the
rotational invariance. There are two methods of the realization of
such scheme. A first one based on the Christ, Friedberg and Lee
(CFL) technique \cite{RL}.



Commonly CFL technique leads into complicated geometrical analysis.
For our task it would be very useful to use the analogy between Random
Lattice and Random Surface technique that was revealed  recently \cite{david, bogacz}.
The idea is quit simple:
for beginning let us consider a lattice with fixed positions (for simplicity it is possible
to use the trivial HC lattice, where basis vectors are just $\vec \nu= \vec i, \vec j \ldots $)
in a flat space. For a simulation of the Random Lattice let us consider
small deformations of the geometry of this space ($\gamma_{ij} \to
g_{ij}$) so that one can rewrite the problem of the
random lattice averaging in the terms of the random deformations
of the geometry of this space
\cite{david}. This is a standard quantum gravity problem for which powerful
methods of the Matrix Theory could be used.

In our problem we discuss the link integrals that depend on the basis vectors
$\nu$. All such integrals are considered separately for any lattice site $x_i$.
It means that rotation invariance violation artifacts could be avoided by
considering only rotation deformations of these basis vectors (translation and
re-scaling deformation are left aside in our case).

Let $R \in \mbox{SO(4)}$ be a rotation operator of 4 dimensional vectors $\nu$
$$
\nu_i' = R_{ij} \nu_j \, .
$$

It is essential to note that our task can be
reformulated in the language of the standard Hermitian averaging because
$\mbox{SO}(4) = \mbox{SU}(2) \times \mbox{SU}(2)$.
For infinitesimal rotations we obviously have
\begin{equation}
R_{ij} = \delta_{ij} + H_{ij} \, ,  \label{infrot}
\end{equation}
where $H_{ij}$ is a traceless antisymmetric matrix.

Consider the Gaussian Ensemble $\hbox{\mygotic H}$ of such arbitrary
rotation. The matrix average of
an arbitrary function $f$ with respect to Gaussian measure is
$$
\langle f \rangle_{\hbox{\mygotic H}} = \frac{1}{N_0} \int d H e^{-
\mbox{tr}(H^2)/2} f( H ) \, .
$$
where $ dH $  is the standard Haar measure.
The normalization factor $N_0$ is fixed by requiring that
$\langle f = 1 \rangle_{\hbox{\mygotic H}} = 1$.

Using of the Matrix integration technique one can prove the so-called
Matrix Wick's theorem for traceless antisymmetric matrices
\cite{Mulase:2002cr, david, RM, bogacz}
$$
\langle H_{ij} H_{kl} \rangle_{\hbox{\mygotic H}} =  \delta_{ij}
\delta_{jl} - \delta_{il} \delta_{jk} ,
$$
\begin{equation}
\langle \prod_{(i,j)} H_{ij} \rangle_{\hbox{\mygotic H}} = \sum_{\mbox{pairings}} \,
(-1)^{\kappa} \prod_{\mbox{pairs}} \langle H_{ij} H_{kl}\rangle_{\hbox{\mygotic H}} \, ,
\label{pair}
\end{equation}
where the sum extends over all possible pairings and $\kappa$ is the number of crossings in the
pairing. The matrix average of any odd combination of $H_{ij}$ equals zero due to the
parity argument.

It is not hard to prove that the main contribution into the averaging sum
over such infinitesimal rotations comes from the original non-deformable lattice
(connected with the $\delta_{ij}$ part in (\ref{infrot})).
In order to cancel non-deformable lattice artifacts let us consider the ensemble
$\hbox{\mygotic H}$ without this non-deformable contribution:
$\hbox{\mygotic H}' = \hbox{\mygotic H}  - \hbox{\mygotic 0} $.
Using such averaging, for leading order of chiral effective lagrangian one
gets
$$
\langle \mbox{tr} (\alpha) \rangle_{\hbox{\mygotic H}'} =
\langle \mbox{tr} (a^2 \partial_\nu U \partial_\nu U^+) \rangle_{\hbox{\mygotic H}'} =
 \mbox{tr} (3 a^2 \partial_i U \partial_i U^+) =
  \mbox{tr} (3 a^2 (L_i L_i) ),
$$
where $L_i=U^+ \partial_i U$,
and for the NL correction one obtains
$$
\langle \mbox{tr} (\alpha^2) \rangle_{\hbox{\mygotic H}'} =
\langle \mbox{tr} (a^4 \partial_\nu U \partial_\nu U^+
                 \partial_\nu U \partial_\nu U^+) \rangle_{\hbox{\mygotic H}'} =
  \mbox{tr} (3^2 a^4 ((L_i L_i)^2 - \frac{1}{2}[L_i,L_j]^2) ) =.
$$
\begin{equation}
= - 3^2 a^4 \left( \frac{1}{2}\mbox{tr}^2(L_i L_i) + \mbox{tr}^2(L_i L_j) -
4 \mbox{tr} (L_i L_i L_j L_j) \right)  =
\label{NLorder}
\end{equation}
These results reproduce the HL and BCHL results and it is easy to see that this
is just what we expected to receive because
this contribution was obtained from many other approaches
\cite{Gasser}.
In such a way, we can apply this procedure to corrections of any orders
from (\ref{comp}) and obtain the rotation invariant result due to the pairing.
It means that the question about the derivation of the chiral effective
lagrangian from Lattice QCD become just a combinatorial task.

It is interesting to point out that our $SU(2)$-flavor result for coefficients in the NL order
contribution of the Chiral Perturbation Theory (\ref{NLorder})
$$
L^r_1=\frac{L^r_2}{2}=-\frac{L^r_3}{4}
$$
is in agreement with experiment data from $\pi\pi \rightarrow \pi\pi$
scattering \cite{Gasser} and these coefficients torn out to be closed
to the model prediction of the $V$ exchange \cite{Pich}.

In the last part of this section we show an application of our procedure. We
will find that an infinite subseries of the chiral perturbation can be
summed up into the Born-Infeld form.

If the expression (\ref{pair}) allows us to calculate all
terms in expansion (\ref{comp}), let us consider just the first
column there. It is easy to show that
either of these is proportional to some power of the leading order
contribution (\ref{lorder})
\begin{equation}
\begin{array}{ccccc}
\mbox{tr}[\alpha] &\sim& \mbox{tr} [ L_\mu L^\mu ]& + & \cdots \\
\mbox{tr}[\alpha^2] &\sim& \mbox{tr} [ (L_\mu L^\mu )^2 ]& + & \cdots \\
\mbox{tr}[\alpha^3] &\sim& \mbox{tr} [ (L_\mu L^\mu )^3 ]& + & \cdots \\
\cdots\cdots & \cdots & \cdots\cdots & \cdots & \cdots\cdots \\
\mbox{tr}[\alpha^n] &\sim& \mbox{tr} [ (L_\mu L^\mu)^n ]& + &
\cdots\\
\cdots\cdots & \cdots & \cdots\cdots & \cdots & \cdots\cdots
\end{array}
\label{ser}
\end{equation}
Substituting (\ref{ser}) into (\ref{seff}) and collecting  all
terms which depend on the power of the prototype lagrangian one
obtains the following expression for the effective chiral lagrangian
\begin{equation}
{\cal L}_{\mbox{eff}} \sim - \mbox{tr} \left[ 1 - \sqrt{1-1/\beta^2
L_\mu  L^\mu } \right] - \mbox{tr} \left[ \log(1 -
\frac{1}{2}(1-  \sqrt{1-1/\beta^2
L_\mu L^\mu })) \right] + \cdots ,
\label{otvet}
\end{equation}
where $\cdots$ are all other terms (in particular the Skyrme term)
and $\beta$ is an effective coupling constant that depends on the
value of our stationary point $u_0$.

Now let us discuss the result (\ref{otvet}). It was obtained that
some part of chiral effective action has a Born-Infeld form plus a
first logarithmic correction to it. In \cite{soliton}, it was shown
that such form of the effective action plays a very essential role in
the problem of the chiral bag formation because just these
square-root terms generate the step-like distribution solutions that
can be interpreted as internal phases in the two-phase model of the
low-energy baryon states. Another terms play essential role only on
large distances from the confinement surface and can be considered as
corrections.

\section{Conclusions}
The aim of this paper is to derive the chiral effective lagrangian
from QCD on the lattice at the strong coupling limit. We find that
this theory looks like a Born-Infeld theory for the prototype chiral
lagrangian. Such form of the effective lagrangian is expected. From
the methodological point of view our consideration is very similar
to the low-energy theorem in string theory that leads to the
Born-Infeld action \cite{IIB}. Moreover, in \cite{soliton}, it was
shown that Chiral Born-Infeld Theory (without logarithmic
corrections) has very interesting   ``bag"-like solutions for chiral
fields. It was an additional motivation of our work.

The Chiral Born-Infeld theory is a good candidate for the role of the
effective chiral theory and a model for a chiral cloud of
baryons. In this model one can find not only spherical ``bags", it is
possible also to study the ``string"-like, toroidal or
``\textbf{Y}-Sign"-like solutions, or some other geometry. The
geometry of the confinement surface  depends directly on the model
of color confinement and it would be very interesting to use, for
example, the Lattice QCD simulations for the color degrees of
freedom in combination with our model for the external chiral field.

\vspace{1cm}

This work is partially supported by the Russian Federation
President's Grant 1450-2003-2. The hospitality and financial support
of the ECT* in Trento is gratefully acknowledged.

\end{document}